# Dimethylsilanone generation from pyrolysis of polysiloxanes filled with nanosized silica and ceria/silica

Kostiantyn Kulyk,*[a] Henning Zettergren,[a] Michael Gatchell,[a] John D. Alexander,[a] Mykola Borysenko,[b] Borys Palianytsia,[b] Mats Larsson,[a] and Tetiana Kulik[b]

[a] Dr. K. Kulyk, Prof. H. Zettergren, Dr. M. Gatchell, Dr. J.D. Alexander, Prof. M. Larsson
Department of Physics
Stockholm University
SE-106 91, Stockholm, Sweden
E-mail: kostiantyn.kulyk@fysik.su.se

[b] Dr. M. Borysenko, B. Palianytsia, Dr. T. Kulik
Chuiko Institute of Surface Chemistry
National Academy of Sciences of Ukraine
17 General Naumov Street, 03164 Kyiv, Ukraine

**Abstract:** Polydimethylsiloxane (PDMS) is a widely used organosilicon polymer often employed in formulations with fine oxide particles for various high temperature applications. Although PDMS is considered to be thermally stable and chemically inert, it is not always clear how the oxide filler influences its thermoresistance, decomposition chemistry and what reactive products are formed in the underlying thermal reactions. In this work we use temperature programmed desorption mass spectrometry (TPD MS) to study the pyrolysis of PDMS and its composites with nanosized silica and ceria/silica. Our results suggest that the elusive organosilicon compound – dimethylsilanone is generated from PDMS over a broad temperature range (in some cases starting at 70 °C). The presence of nano-oxides catalyzed this process. Ions characteristic of the fragmentation of dimethylsilanone under electron ionization were assigned with the aid of DFT structure calculations. Possible reaction mechanisms for generating dimethylsilanone were discussed in the context of the calculated kinetic parameters. Observed accompanying products of PDMS pyrolysis, such as tetramethylcyclodisiloxane and hexamethylcyclotrisiloxane, indicate that multiple channels are involved in the dimethylsilanone release.

## Introduction

Extensive commercial utilization of polydimethylsiloxane (PDMS) polymer has been driving intense investigations of its composites upgraded by nanosized inorganic fillers. The introduction of small amount of transition and/or rare-earth metal nano-oxides is known to induce significant improvements in the specific properties of the polymer, such as thermal stability, conductivity, texture, hydrophobicity, interfacial behavior, etc.[1-7] PDMS/silica is an example of such a hybrid nanocomposite material that is already manufactured and applied at a multi-tonnage scale.

A considerable part of PDMS/silica compositions is utilized in fields where thermal stability and resistance to oxidation is a necessity. For example, in food-related processes PDMS/silica is primarily used as an anticaking agent in confectionery and flour products, and an antifoaming agent in edible oils.[8] These applications account for about 7% of the total usage of PDMS, the most widely used organosilicon polymer.[9] PDMS is included in the current version of the general standard for food additives (GSFA) codex for use in a wide range of edibles at acceptable maximum levels of 10-100 mg/kg food,[8] among which are vegetable oils and fats – products that are generally designated for cooking and frying. In food processing, straight-chain PDMS is typically applied in formulations with fumed silica because such mixtures are considered to be more effective foam control agents than the individual components.[10] The development of modern defoamers continues to use PDMS/silica combinations in order to adjust for different types of foam related issues in industry.[9-11]

Knowing that the decomposition chemistry and thermal stability of the polymer can be altered by the introduction of highly dispersed fillers Kulyk et al. recently studied the thermally induced processes in PDMS/silica and PDMS/silica/ceria nanocomposites by using a combination of temperature programmed desorption mass spectrometry (TPD MS) and thermogravimetric analysis (TGA).[12] These results showed that the thermal degradation of PDMS adsorbed on $SiO_2$ and $CeO_2/SiO_2$ nano-oxides proceed via three main stages: 1) covalent attachment of PDMS to the silica via the electrophilic substitution of a surface silanol by a terminal trimethylsilyl group of the polymer; 2) formation of the cyclic oligomer – hexamethylcyclotrisiloxane (HMCTS); 3) high temperature degradation of the polymer accompanied by the formation of methane and ethylene. The observed PDMS depolymerization and radical degradation processes were in general found to be in a good agreement with those reported earlier.[2,13-15]

However, there is a controversy in the literature regarding the high temperature utilization of PDMS. On the one side many works, in particular those dealing with polymer's industrial applications, suggest that polydimethylsiloxane-based materials are well-suited for foam control in food-processing due to their thermal stability and chemical inertness.[9,10] On the other side, various linear and cyclic PDMS were reported to thermally generate dimethylsilanone, $(CH_3)_2Si=O$, the silicon analogue of acetone – a compound which is believed to be extremely reactive.[15-21] Indeed, polydimethylsiloxane-based silicon greases, that are believed to be chemically "inactive entities", have been the subject of two review articles.[22,23] These reviews indicated that

PDMS can actually function as a source of $(CH_3)_2Si=O$, leading to undesirable side reactions that yield various exotic molecular and supramolecular compounds.[22,23] In this work we aim to shed further light on these intriguing issues by studying dimethylsilanone generation by PDMS and its composites with nanosized silica and ceria/silica.

Up until recently such multiply bonded compounds of silicon were unknown at ambient conditions and were considered to be elusive species. But presently the chemistry of silicon is undergoing rapid development resulting from major breakthroughs that have been achieved in the particular field of the silicon ketone analogues – silanones ($R_2Si=O$). Firstly, the remarkable technique devised by Xiong et al. to partially stabilize the silicon-oxygen double bond by coordination of Lewis bases to the Si atom allowed synthesis of several compounds that feature the given structural fragment.[24-25] Secondly, pioneering work by Filippou et al. resulted in the synthesis and isolation of the first stable genuine silanone.[26] Finally, Wang et al. successfully stabilized and isolated monomeric silicon oxides $Si_2O_3$ and $Si_2O_4$ for the first time.[27]

Prior to this progress in the preparation of stabilized compounds with Si=O bonds, isolable silanones were unknown at ambient conditions due to their high reactivity. Previously they had only been detected and spectroscopically studied in solid noble gas matrices at low temperatures.[28-31] The existence of the silanone molecules and ions were also observed in the gas phase.[32-33]

Several experimental studies have suggested that silanones are formed as intermediates in the thermal reactions of low-molecular organosilicon compounds such as silylperoxides and hydrosilylperoxides, siloxetanes, silenes, hydridosilylketenes, allyloxysilanes, alkoxyvinylsilanes and polysilylated diazomethanes.[16,34-39] Linear and cyclic PDMS were reported to generate dimethylsilanone.[15-21] These claims of dimethylsilanone formation were usually based on kinetic data and chemical trapping experiments.[16-18] Among the direct physical methods, matrix isolation IR spectroscopy was used to identify this compound being involved as an intermediate in the thermal redistribution of some low-molecular cyclic and linear siloxanes.[28-31] Mass spectrometric detection of dimethylsilanone as a product of thermal degradation in polydimethylsiloxane/montmorillonite nanocomposites, however, has only been reported once.[15]

The lack of mass spectrometry (MS) data and especially temperature programmed desorption MS studies is surprising given the broad utilization of siloxane/oxide composites and the information about reaction kinetics that can be obtained by this technique.[40-42] In the present study, we have for the first time utilized the capabilities of TPD MS to study pyrolysis of pure PDMS and its compositions with silica and ceria/silica focusing on dimethylsilanone release. Here, we report the detection of dimethylsilanone generation during the pyrolysis of pure PDMS, PDMS/silica and PDMS/silica/ceria nanocomposites over a broad temperature range. The experimental results are discussed in view of molecular structures and dissociation energies from density functional theory (DFT) calculations. The detection of this reactive compound being generated over a broad temperature range (up to 700 °C) is important for multiple high temperature applications of PDMS/silica nanocomposites.

Another important aspect of this work is related to the growth of environmental contamination by siloxanes and the emerging issues associated with it. Sixty years of intensive utilization of polysiloxanes has led to their widespread distribution in the environment. Nowadays various siloxanes can be found in soil, air, and water.[43,44] While the health effects of increasing concentrations of organosilicon environmental pollutants on humans and animals are yet to be understood, their harmful impact on renewable energy production is already considered to be a great problem.[44-46] For example, steadily increasing amounts of siloxanes in landfills and sewage sludge, and consequently in the biogas derived from those sources, creates new technical challenges. In particular, combustion of siloxane-contaminated biogas produces abrasive microcrystalline silica that causes serious damage to gas engines, heat exchangers and catalytic exhaust gas treatment systems.[47]

Methods of purifying biogas therefore have huge commercial importance since biogas production is one of the potential avenues being considered by countries to help them reach their renewable energy targets.[44,46] Removal of volatile polysiloxane pollutants from biogas by high surface area adsorbents, including silica-gels, zeolites and mixed alumina/silica systems, is the most efficient technology currently used.[44,47-49] However, cost-efficient methods of regenerating the adsorbents, which is conventionally performed by thermal desorption of adsorbed pollutants, are yet to be found.[50,51] Due to the complexity of the thermal desorption process and chemical reactions involved, the factors responsible for the significantly lowered adsorption capacity of regenerated adsorbents are still not fully understood.[50-52] The temperature programmed desorption mass spectrometry data in this article gives information on the thermal profile of dimethylsilanone release from adsorbed PDMS under heating in vacuo. This observation advances the current understanding of polysiloxane thermal degradation and accompanying reactions, including the previously reported surface catalyzed depolymerization and chemisorption (see also Ref. 12), which may play important role in adsorbent poisoning.

**Experimental Section**

**Temperature programmed desorption mass spectrometry (TPD MS)**

The TPD MS experiments were carried out using a monopole mass spectrometer MKh-7304A (Electron, Sumy, Ukraine) with electron ionization. The experimental procedure implemented here was the same as in a recent study of Kulyk et al.[12] Briefly, the measurements were performed as follows: a 15 mg of oxide/polymer sample was put into the mass spectrometer ampoule and evacuated to $5 \cdot 10^{-5}$ Pa. It was then heated from room temperature up to 800 °C at a controlled rate of 10 °C/min. Throughout the heating process desorbed species were ionized and mass analyzed giving a desorption profile of the sample as a function of temperature. All desorption peaks were recorded. Full details of this TPD MS setup, equipment used and methods of kinetics determination are given elsewhere.[12,41,42]

**Preparation of polymer/oxide nanocomposites**

Fumed silica A-300 (specific surface area ($S_{Ar}$) = 319 $m^2/g$; primary particle size = 8 nm in diameter) was supplied by a pilot plant of the Chuiko Institute of Surface Chemistry (Kalush, Ukraine). Cerium dioxide-containing mixed nano-oxides $CeO_2/SiO_2$ were prepared using cerium (III) acetylacetonate hydrate (Sigma-Aldrich) and the above mentioned silica via the procedure reported in detail elsewhere.[12] Two $CeO_2$-containing samples were used – $CeO_2/SiO_2$_low (6.6 wt. % of $CeO_2$; $S_{Ar}$ = 265 $m^2/g$; consisting of amorphous nanoceria according to XRD) and $CeO_2/SiO_2$_high (23.3 wt. % of $CeO_2$; $S_{Ar}$ = 189 $m^2/g$; consisting of 3 nm crystalline ceria in a cubic lattice arrangement according to XRD data). Polydimethylsiloxane − PDMS-1000 (molecular weight Wm ≈ 7960, degree of polymerization dp = 105; referred to here as PDMS) was purchased from Kremniypolymer (Zaporizhya, Ukraine). Preparation of PDMS/oxide composites was done by adsorption of the polymer onto the oxide samples as follows: the samples of $CeO_2/SiO_2$ and fumed silica were dried at 550 °C for 1 hour. Solutions of PDMS in hexane were then mixed with the oxide samples, stirred for 30 minutes and dried at room temperature. After evaporation of the solvent, the nominal composition of all $PDMS/CeO_2/SiO_2$ samples was: 40/60 % wt. polymer/nano-oxide.

**Results and Discussion**

While the thermal degradation pattern of simple polysiloxane systems is relatively well studied and understood, the identification of chemical reactions occurring during the decomposition of polymers filled with nanoparticles is more complicated. For example, surface assisted reactions can change the thermal decomposition chemistry considerably. To gain insight into the role of dimethylsilanone in PDMS pyrolysis the underlying processes need to be unraveled.

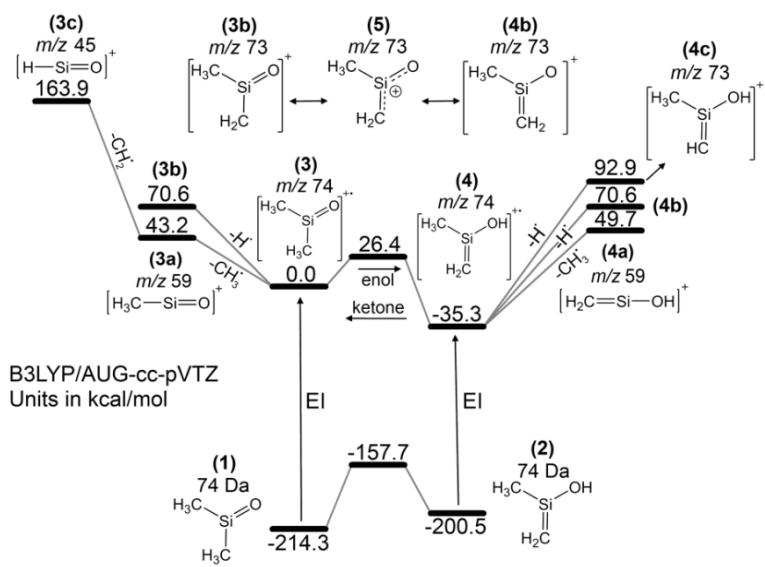

**Scheme 1.** Calculated molecular structures, transition state energies, and dissociation energies for neutral dimethylsilanone as well as for its cationic molecular and fragment ions (the values of the m/z ratios for the ions detected by MS are placed above the corresponding structures). We used dimethylsilanone ketone cation (3) as our reference ion with 0.0 kcal/mol.

Earlier we reported that the thermal degradation of PDMS adsorbed on silica and nano-oxides $CeO_2/SiO_2$ proceeds via three main stages:[12] 1) covalent attachment of PDMS to the silica via the electrophilic substitution of a surface silanol by a terminal trimethylsilyl group of the polymer; as a result of this reaction, trimethylsilanol is released (characteristic fragment ion is *m/z* = 75); 2) formation of the cyclic oligomer – hexamethylcyclotrisiloxane (HMCTS, characteristic fragment ion is *m/z* = 207); 3) high temperature degradation of the polymer accompanied by the formation of methane and ethylene (characteristic fragment ions are *m/z* = 16 and 28, correspondingly). Cerium dioxide nanoparticles were found to catalyze PDMS depolymerization as well as the generation of methane and ethylene.

Among the above mentioned reactions, the formation and desorption of HMCTS should be taken in account while discussing dimethylsilanone generation. Dimethylsilanone has been claimed to be involved in a variety of thermal redistribution reactions of HMCTS and related cyclic polymethylsiloxanes.[17-19,53]

**Detection and fragmentation of dimethylsilanone**

The high polarity of the silicon-oxygen double bond as well as the diffuse nature and relative weakness of the corresponding π-bond are responsible for the high reactivity of dimethylsilanone. The compound is believed to undergo facile dimerization and oligomerization, which is the main reason for it being difficult to isolate and detect. However, due to the specific conditions of TPD MS experiments, such as fast pumping of the volatile products of the surface pyrolytic reactions under high vacuum and lack of oxygen, acquisition of ions corresponding to the fragments of eliminated and desorbed dimethylsilanone was possible. Under those TPD MS conditions dimethylsilanone was not able to participate in addition and condensation reactions before entering the ion source, where it was ionized by electrons and fragmented.

To aid in the interpretation of the experimental results we have calculated the molecular structures and dissociation energies for dimethylsilanone using density functional theory (DFT) as implemented in the GAUSSIAN09 software.[54] For the neutral dimethylsilanone molecule, the ground state structure is of keto-form (1, Scheme 1). The enol form of the neutral molecule has a total binding energy (enthalpy) that is 13.8 kcal/mol higher at B3LYP/Aug-CC-pVTZ level of theory (2, Scheme 1). The isomerization from the keto (1) to the enol (2) form has a barrier of 56.6 kcal/mol (Scheme 1), which indicates that the formation of a ketone is strongly favored at chemical equilibrium. For this reason it is likely that dimethylsilanone formed in studied pyrolytic reactions at the surfaces was desorbing as a ketone.

The electron ionization mass spectrum of the sample of pure PDMS recorded at 555 °C is shown in Figure 1. In this temperature range there is little interference between ions related to dimethylsilanone desorption and to desorption of the rest of the products. The intact dimethylsilanone ion with $m/z$ = 74 is clearly visible together with its fragmentation products.

For the cations the relationship between total binding energies at B3LYP/Aug-CC-pVTZ level of theory is reversed relative to the neutral dimethylsilanone tautomers, with the enol structure (4, Scheme 1) being energetically favorable by 35.3 kcal/mol compared to the ketone structure (3, Scheme 1). The isomerization of the cation from ketone to enol structure is however slowed by a 26.4 kcal/mol barrier.

The most prominent peak corresponding to $m/z$ = 73 in the mass spectrum (3b, Scheme 1; Figure 1) is about five times stronger than the peak for the intact dimethylsilanone ion. Molecular ion 3 may fragment through: 1) loss of a hydrogen atom leading to the formation of cation 3b with $m/z$ = 73; 2) loss of a methyl radical, which leads to cation 3a with $m/z$ = 59. Between these two primary fragment ions, ion 3b is more abundant in the mass spectrum (Figure 1).

The loss of the methyl radical is the result of homolytic cleavage of Si-C bond leading to the formation of the silicon analogue of the acylium ion (3a) with $m/z$ = 59 (Scheme 1; Figure 1). The present calculations show that the dissociation energy for Si-C bond cleavage is 43.2 kcal/mol, which is lower than for C-H bond cleavage (70.6 kcal/mol) leading to the formation of ion 3b. One possible explanation for the high abundance of ion 3b is that the frequency of the C-H bond vibration is higher than that for the Si-C bond. Using a simple Arrhenius type relationship, C-H bond cleavage is then expected to be associated with a higher pre-exponential factor. This may compensate for the somewhat higher dissociation energy and lead to a higher decay rate. Another possibility is ultrafast emission (sub-picosecond timescale) of an H atom from the electronically excited dimethylsilanone cation.[55]

The fragment internal energy is in many cases sufficient to initiate further decay processes, e.g. to form ion 3c with $m/z$ = 45 (~ 25 %) by removing $CH_2$ from the methyl group of 3a leaving behind only a hydrogen atom. The decay of ion 3b also leads to the formation of ions 3a and 3c by the removal of $CH_3$ groups followed by H transfer to the residual ion, see transitions from 3b to 3a, and from 3a to 3c in Scheme 1. H transfer from a leaving $CH_3$ group is a known fragmentation mechanism occurring under the electron ionization of silicon-organic compounds.[56,57]

The assignment of ions 3 and 3b to the fragmentation of dimethylsilanone and ions 3a and 3c to the products of its decay, and not to the decay of other ions, is consistent with the TPD MS data presented below. It is observed that the TPD curves corresponding to these ions follow the same trend, which indicates that they result from a single type of desorption product.

The right part of the energy diagram in Scheme 1 is related to the dissociation pathways following dimethylsilanone enolization. Conversion of dimethylsilanone to its tautomer – methyl(methylene)silanol (4) is associated with an energy barrier of 26.4 kcal/mol, which is significantly lower than the lowest dissociation energy and thus expected to occur under the present experimental conditions. Ions that are likely to be formed as a result of fragmentation of this ion (4) are shown as 4a, 4b, and 4c in Scheme 1. Note that 4b has the same energy as 3b in the left part of the diagram, the reason being that they represent two resonant structures of the same ion (5), as indicated in the upper part of Scheme 1. Ion 4c may also contribute to the peak with $m/z$ = 73 observed in TPD MS measurements, but it requires significantly higher energy and it is thus not expected to be a prominent decay pathway. Interestingly, the Si-C bond cleavage is energetically less favorable from the enol compared to ketone structure of dimethylsilanone (Scheme 1). Thus the overall decay rate for Si-C bond cleavage is expected to be reduced. As a consequence, the branching ratio for C-H bond cleavage will be larger than if only the ketone structure would have been populated.

In this context, it is worth mentioning that the ion with $m/z$ = 73 is sometimes interpreted as the $(CH_3)_3Si^+$ cation in the literature.[44,56-58] It is believed that it can be present in the fragmentation of silicone compounds with trimethylsilyl groups in their structure. For example, this is the most abundant ion in the electron ionization mass spectrum of tetramethylsilane[59] and for trimethylsilane it has an abundance of 45% relative to the main fragment ion with $m/z$ = 59. Some studies[20,21] suggest the possibility of forming trimethylsilane and tetramethylsilane during the pyrolysis of polydimethylsiloxanes. This raises the question of whether the observed ion with *m/z* = 73 is really attributed to dimethylsilanone, or if it results from fragmentation of tetra- and trimethylsiloxanes formed from the PDMS as it is being heated. In the present work we observe this ion at relatively low temperatures for some samples. However, the radical reaction mechanism required for methyl group migrations to occur (a possible path to permethylated silanes), is known to require higher temperatures.[12,14] This suggests that the formation of dimethylsilanone is much more likely than the formation of permethylated silanes under the present experimental conditions. Indeed, it was found earlier that the trimethylsilyl terminal groups of PDMS are participating in the chemisorption of the polymer onto the surface of nano-silica. As a result of the reaction trimethylsilyl moieties are lost in the form of trimethylsilanol which desorbs from the surface. A detailed discussion about possible reaction pathways is presented in the section below. Moreover, the ion with *m/z* = 73 was recently interpreted as dimethylsilanone in the enolate form in other pyrolytic studies.[15] The doubts concerning the nature of the origin of dimethylsilanone ion in our spectra can be eliminated by carrying out comparative analysis of the relative intensities of the major fragments in the spectra of tetramethylsilane, trimethylsilane and our TPD MS spectra.[59]

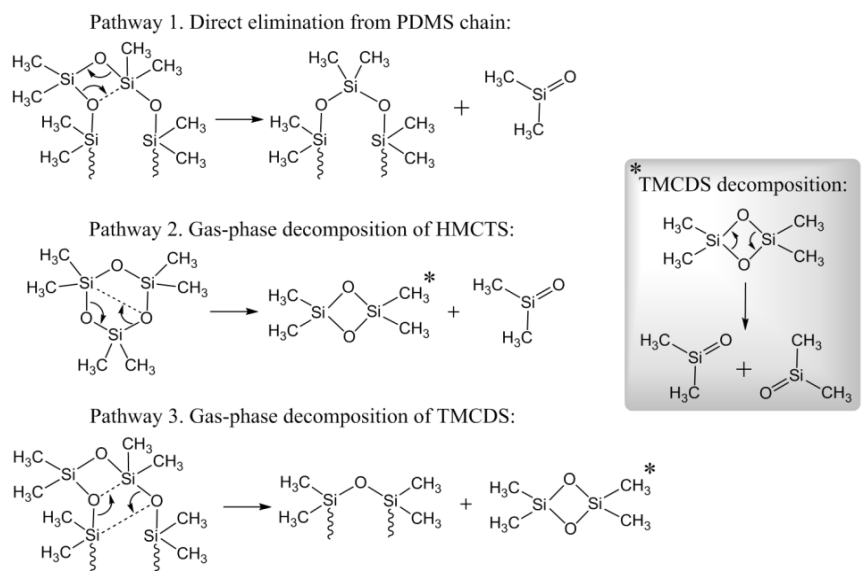

**Scheme 2.** Possible formation routes of the observed dimethylsilanone.

According to the NIST database[59] and work by Basner et al.,[57] in the electron ionization spectrum of tetramethylsilane the following fragment pattern is present: *m/z* = 73 (100 %), 74 (10 %), 75 (~ 3 %) 59 (2-3 %), 45 (10 %), 43 (~ 16 %). The spectrum of trimethylsilane is characterized by the fragments: *m/z* = 59 (100 %), 73 (45 %), 74 (2-3 %), 75 (~ 1-2 %), 45 (10 %), 43 (~ 30 %). We observed a different ratio for the dimethylsilanone fragment ion intensities – *m/z* = 73 (100 %), 74 (~ 16 %), 75 (~ 9 %), 76 (~ 1.5 %), 59 (~ 9 %), 45 (~ 25 %), see Fig. 1. It should be mentioned that these ratios depend on the ion temperature, which may vary in different experiments. However, we believe that it is safe to attribute this electron ionization fragmentation pattern to the fragmentation of dimethylsilanone because at this temperature (555 °C) little generation of other products was observed and all of them were identified. The peaks with smaller *m/z* = 28 and 16 originate from ethylene and methane formed as a result of high temperature radical degradation of the PDMS – a process that has no direct relation to the formation of dimethylsilanone[12]. The peaks with higher *m/z* = 207, 191, 133 and 96 originate from fragmentation of HMCDS.[12] The peak with *m/z* = 147 corresponds to the tetramethylcyclodisiloxane cation (TMCDS), its role is discussed further. Consequently, the possibility of contribution from fragment ions resulting from the decay of heavier organosilicon species is unlikely.

Notably, we observe the isotope peak M+2 (*m/z* 76) in the spectrum (Figure 1). It is another indication that we detect dimethylsilanone as both possible contaminants trimethylsilane and tetramethylsilane do not exhibit the ion with *m/z* 76 in their fragmentation patterns.[59] Of course, the molecular ion of trimethylsilane $(CH_3)_3SiH$ would have exhibited the M+2 isotope peak (*m/z* 76) with an abundance of 3.6 %, but the trimethylsilane molecular ion M (*m/z* 74) itself represents only 2-3 % of the intensity of its main fragment *m/z* = 59. Thus the expected M+2 isotope peak intensity of trimethylsilane would be equal to 0.07-0.11 % relative to the abundance of *m/z* 59, which makes it undetectable.

To further support our interpretation we compare the shape of the TPD curves for different fragments. This makes it possible to determine whether the fragments originate from a common parent ion. In the current results the TPD profiles of the peaks that correspond to dimethylsilanone generation (*m/z* = 74, 73, 59, 45) display very similar shapes and temperature maxima. Specifically, $T_{max}$ for the dimethylsilanone desorption from pure PDMS, Fig. 2, is 550.3 ± 1.6 °C. For comparison, $T_{max}$ for the HMCTS desorption of is ~ 525 °C (Fig. 2, Table 1). The temperature of the maximum reaction rate, which in the case of TPD MS corresponds to the ion intensity, was also extracted from the TPD MS data. In a similar manner to that used by many authors,[60] an approximate calculation of the activation energy of the thermal processes was made. The calculation of the kinetic parameters of the integrated intensities of the TPD peaks for ions with *m/z* = 74, 73, 59 and 45 gave a similar activation energy and a pre-exponential factor of the same order of magnitude (~ $10^6$ $s^{-1}$).This strongly suggests that these ions were indeed dimethylsilanone fragment ions.

**Dimethylsilanone formation and desorption**

The dimethylsilanone elimination from a polydimethylsiloxane chain is illustrated in Scheme 2 (Pathway 1). The mechanism proposed is consistent with the general conclusion made in the comprehensive review by Voronkov describing the pyrolysis of both linear and cyclic siloxanes as proceeding via dimethylsilanone elimination through a four-centered transition state.[19]

According to Scheme 2 (Pathway 1) elimination of dimethylsilanone occurs when the polymer chain takes a conformation suitable for the formation of the four-centered transition state. When this happens the oxygen atom acts as a intramolecular nucleophile attacking the silicon atom located next to the nearest [–SiO(CH$_3$)$_2$–] monomeric unit. In the process, the two electrons creating a bond between the attacked silicon and the other oxygen are repelled towards the oxygen. This enables them to further participate in the formation of the silicon-oxygen double bond of the leaving molecule (Scheme 1, Pathway 1). The formation of this new silicon-oxygen bond is immediately followed by the elimination of the dimethylsilanone molecule and polymer chain shortening (Scheme 2, Pathway 1).

In the following sections, we show that multiple channels are involved in the process of dimethylsilanone release. This is concluded from an analysis of the TPD MS profiles of various ions for different samples and observation of the pyrolysis products (namely – HMCTS and tetramethylcyclodisiloxane (TMCDS)) that may be participating in its gas-phase generation. These two alternative channels are depicted in Scheme 2 (Pathways 2 and 3).

**Pyrolysis of pure PDMS**

Fig. 2 shows the TPD MS curves obtained during the pyrolysis of a pure polymer sample. Numbers above the curves correspond to the *m/z* values together with the supposed structures of the ions.

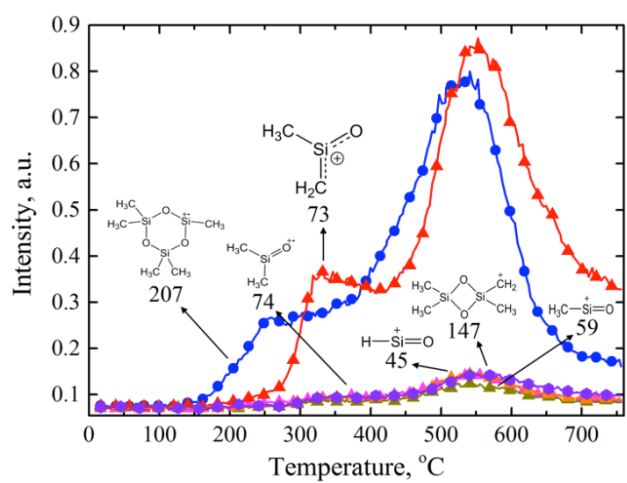

**Figure 2.** TPD MS curves for a pure PDMS sample. Numbers above the curves correspond to the *m/z* values – drawings to the supposed structures of ions.

On the basis of previous work, the ion with *m/z* = 207 can be attributed to the characteristic fragment of HMCTS – pentamethylcyclotrisiloxane ion radical (depicted in Fig. 2).[12] HMCTS desorption starts at 140 °C and continues up to 750 °C (Figure 2). The formation of dimethylsilanone by gas-phase decomposition of HMCTS is possible by the reaction depicted in Scheme 2 (Pathway 2). Such a possibility has earlier been discussed by Voronkov.[19] However, fragment ions (m/z = 207 and 73) corresponding to these two products have different TPD profiles which indicates that they represent two distinct processes, at least in some temperature regions.

Interestingly, the intact dimethylsilanone ion (m/z = 74) as well as its main fragment ion with m/z = 73, were observed during the TPD MS analysis of the pure polymer. This is in contrast to the results reported by Lewicki et al.,[15] where the fragment with m/z = 73 (the molecular ion of dimethylsilanone with m/z = 74 was not discussed in that publication) was observed only for polymer samples filled with an organically modified montmorillonite clay. The TPD curve for this ion starts to grow at a higher temperature of around 250 °C (Figure 2). It is very important to note that the ions with m/z = 73 and 207 don't exhibit the same trend and have different maxima. It indicates that the processes of formation of HMCTS and dimethylsilanone during PDMS thermal degradation are two independent processes driven by different mechanisms rather than just different stages of one process of the thermal redistribution of cyclic polysiloxanes. This is also supported by the distinct values of the kinetic parameters obtained for these two processes (see Table 1). In contrast, the ions with m/z = 59 and 45, corresponding to the products of further decay of the dimethylsilanone cation (ions 3 and 4 in Scheme 1), follow the same trend as the ion with m/z = 73 (Fig. 2) and have similar values for their kinetic parameters (Table 1). This strongly suggests that all these species (m/z = 74, 73, 59, 45) stem from the same parent molecules, namely dimethylsilanone (see Scheme 1).

Another product that along with HMCTS can be responsible for an alternative route of dimethylsilanone generation is TMCDS. It is known that dimerization of dimethylsilanone to form TMCDS and decomposition of TMCDS to give two dimethylsilanone molecules represent a reversible reaction.[19] Given the high-vacuum nature of the TPD MS experiment it is reasonable to assume that in our case equilibrium shifts towards decomposition of TMCDS according to the Le Chatelier principle (see the process labelled * in Scheme 2). However, ions with m/z = 147 (corresponding to the tetramethylcyclodisiloxane cation) have been observed as a weak signal in our measurements (Fig. 2). This fragile ion appears at different temperatures for pure and filled PDMS pyrolysis. Thus, our TPD MS data shows that in the temperature range 150-250 °C HMCTS is thermally stable, but from 250 °C some of its molecules start to decompose, forming dimethylsilanone and tetramethylcyclodisiloxane, see Fig. 2 and Scheme 2 (Pathway 2).

One would assume that dimethylsilanone and TMCDS are formed due to the decomposition of HMCTS, but already at 340 °C the process of dimethylsilanone formation prevails over the process of formation of HMCTS (Fig. 2), i.e. the dimethylsilanone as well as the TMCDS formations can take place independently from the formation of HMCTS. For example, TMCDS can be directly eliminated from the polymer chain via the reaction illustrated in Scheme 2, Pathway 3. It is also known that TMCDS is a very unstable compound, which after formation immediately splits into two dimethylsilanone molecules.

| Desorption product | m/z | $T_{max}$, °C | n | $E^{\neq}$, kJ·mol$^{-1}$ | $v_0$, sec$^{-1}$, (n=1); l·mol$^{-1}$sec$^{-1}$, (n=2) | $dS^{\neq}$, cal·K$^{-1}$·mol$^{-1}$ | ±S [a], % | $R^2$ [b] |
|---|---|---|---|---|---|---|---|---|
| **PDMS** | | | | | | | | |
| (CH$_3$)$_2$Si=O | 73 | 330 | - | - | - | - | - | - |
| | 73 | 553 | 1 | 140 | 1.85·10$^6$ | -32 | 4 | 0.969 |
| | 73 | 553 | 2 | 265 | 7.31·10$^{14}$ | 8 | 12 | 0.929 |
| | 74 | 549 | 1 | 138 | 1.63·10$^6$ | -32 | 3 | 0.975 |
| | 59 | 550 | 1 | 137 | 1.53·10$^6$ | -32 | 4 | 0.953 |
| | 45 | 549 | 1 | 137 | 1.62·10$^6$ | -32 | 2 | 0.970 |
| TMCDS | 147 | 330 | - | - | - | - | - | - |
| | 147 | 560 | 1 | 122 | 1.25·10$^5$ | -37 | 3 | 0.955 |
| | 147 | 560 | 2 | 272 | 9.22·10$^{14}$ | 8 | 14 | 0.912 |
| HMCTS | 207 | 260 | - | - | - | - | - | - |
| | 207 | 525 | 1 | 102 | 1.16·10$^4$ | -42 | 4 | 0.969 |
| | 207 | 525 | 2 | 208 | 4.34·10$^{11}$ | -7 | 46 | 0.862 |
| **PDMS/SiO$_2$** | | | | | | | | |
| (CH$_3$)$_2$Si=O [c] | 74 | 65 | 1 | 75 | 2.73·10$^9$ | -16 | 4 | 0.978 |
| | 74 | 65 | 2 | 151 | 4.90·10$^{21}$ | 40 | 23 | 0.967 |
| | 73 | 68 | 1 | 77 | 5.28·10$^9$ | -14 | 2 | 0.989 |
| | 73 | 68 | 2 | 137 | 2.86·10$^{19}$ | 31 | 12 | 0.964 |
| | 59 | 71 | 1 | 81 | 1.88·10$^{10}$ | -12 | 2 | 0.987 |
| | 59 | 71 | 2 | 157 | 2.13·10$^{22}$ | 43 | 10 | 0.962 |
| TMCDS [c] | 147 | 74 | 1 | 64 | 2.82·10$^7$ | -25 | 4 | 0.958 |
| | 147 | 74 | 2 | 126 | 1.80·10$^{17}$ | 20 | 5 | 0.974 |
| (CH$_3$)$_2$Si=O | 73 | 645 | - | - | - | - | - | - |
| **PDMS/CeO$_2$/SiO$_2$_low** | | | | | | | | |
| (CH$_3$)$_2$Si=O | 73 | 635 | 1 | 154 | 3.77·10$^6$ | -31 | 4 | 0.967 |
| | 73 | 635 | 2 | 275 | 1.78·10$^{14}$ | 5 | 33 | 0.851 |
| **PDMS/CeO$_2$/SiO$_2$_high** | | | | | | | | |
| (CH$_3$)$_2$Si=O | 73 | 637 | 1 | 164 | 1.26·10$^7$ | -28 | 3 | 0.966 |
| | 73 | 637 | 2 | 287 | 6.67·10$^{14}$ | 7 | 11 | 0.890 |

**Table 1.** Kinetic parameters (temperature of the maximum desorption rate $T_{max}$, reaction order $n$, activation energy $E^{\neq}$, pre-exponential factor $v_0$ and activation entropy $dS^{\neq}$) of the chemical reactions of nanoscale silica and CeO$_2$/SiO$_2$ in compositions with PDMS, where: [a]standard error of the regression; [b]squared coefficient of determination; [c]measurement with increased quantity of sample (50 mg).

We can therefore assume that at least three channels of dimethylsilanone formation contribute to its generation: direct generation from PDMS chains (Scheme 2, Pathway 1), generation by gas-phase decomposition of HMCTS (Scheme 2, Pathway 2) and generation from TMCDS that was formed directly from PDMS chains, but not from HMCTS (Scheme 2, Pathway 3). Comparison of the TPD curves for the ions with $m/z$ = 147 and 73 shows that they go asynchronously for some samples (Figures 2-5) – this supports the idea of three alternative channels of dimethylsilanone formation. Calculated kinetic parameters for the TPD curves for the ions with *m/z* = 147 and 73 have distinct values (Table 1). The temperature maximum for the peak of the ion with *m/z* = 147 is 10 °C higher than the $T_{max}$ values for the ions with *m/z* = 74, 73, 59 and 45, whereas the pre-exponential factor is higher for the latter process. Obtained kinetic parameters indicate that both processes most likely proceed via the first order reaction and through a highly ordered transition state.

Thus, we can conclude that all three processes of dimethylsilanone formation can occur independently – i.e. HMCTS, TMCDS and dimethylsilanone may be formed independently of each other directly from the polymer chain, instead of the cyclic siloxanes. At temperatures above 550 °C the process of dimethylsilanone release prevails over the process of HMCTS release. This behavior can be attributed to the decrease in the polymer chain length – the shorter the length the harder it is for the polymer to acquire the conformation needed for the elimination of a six-membered cycle. This is likely the reason for the stronger contribution of TMCDS and dimethylsilanone in the pyrolysis products as temperature increases.

**Pyrolysis of PDMS/SiO₂ nanocomposite**

Fig. 3 shows the TPD MS profile obtained for the polymer/silica sample.

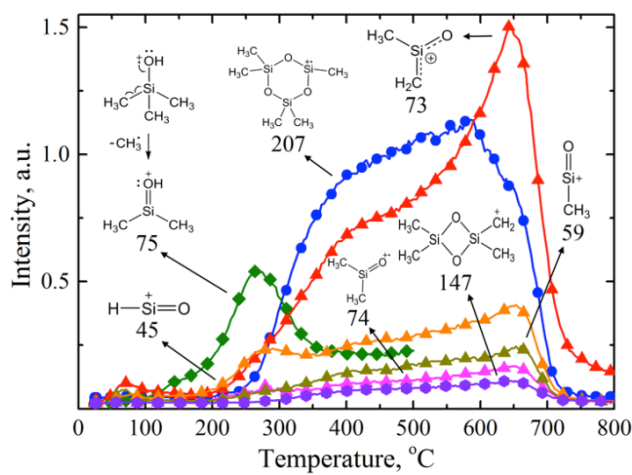

**Figure 3.** TPD MS curves for a PDMS/SiO₂ sample. Numbers above the curves correspond to the *m/z* values.

The figure indicates that silica, which is known to contain weakly acidic surface silanol groups ($pK_a$ = 7.1),[61,62] catalyzes the reaction of dimethylsilanone elimination. A new low-temperature maximum for the ion with *m/z* = 73 (accompanied by its fragments – *m/z* = 59 and 45) appears at 70 °C. The main stage of the dimethylsilanone release, which starts at around 250 °C for the pure polymer, is now lowered by approximately 15 °C. Two more maxima of generated dimethylsilanone (at ~ 450 and 650 °C) are present in the TPD profile of this sample. The TPD curve of the main fragment ion of dimethylsilanone is comparably flatter than that obtained from the pure polymer.

**Pyrolysis of PDMS/CeO$_2$/SiO$_2$ nanocomposites**

Even more dramatic changes in the release of dimethylsilanone were observed for samples containing ceria nanoparticles. For example, the TPD profile of the sample PDMS/CeO$_2$/SiO$_2$_low (containing 6.6 wt. % of CeO$_2$ relative to the total oxide content) is already characterized by the release of dimethylsilanone in a broad temperature range – from 50 °C to 750 °C. A few maxima can be distinguished – at 67, 409 and 634 °C (Fig. 4). Fragment ions with *m/z* = 59 and 45 follow the same trend as their parent (*m/z* = 73). The maximum at 265 °C should be neglected as it is attributed to the release of trimethylsilanol (the ion with *m/z* = 75 is its main fragment ion) and is not related to the dimethylsilanone formation (see Ref. 12 for details).

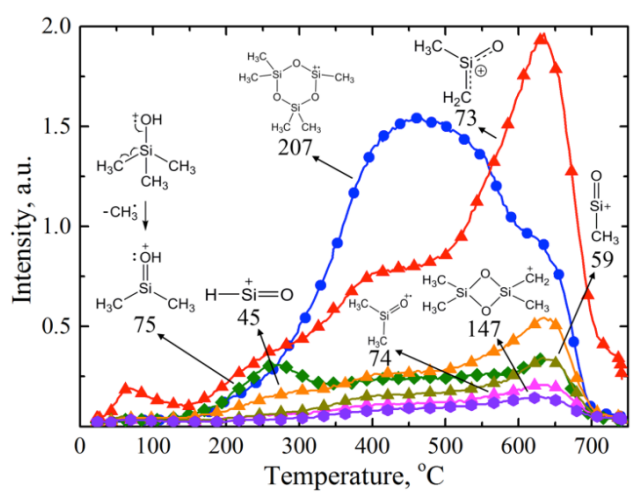

**Figure 4.** TPD MS curves for a PDMS/CeO$_2$/SiO$_2$_low sample. Numbers above the curves correspond to the *m/z* values.

An increase in the content of ceria in the sample (PDMS/CeO$_2$/SiO$_2$_high) was found to even more prominently induce the catalytic action: (CH$_3$)$_2$Si=O generation starts already at 42 °C and continues up to 750 °C, passing through three maxima at 75, 369 and 637 °C in the process (Fig. 5). The maximum at 260 °C can be ignored as it is related to the release of trimethylsilanol.[12]

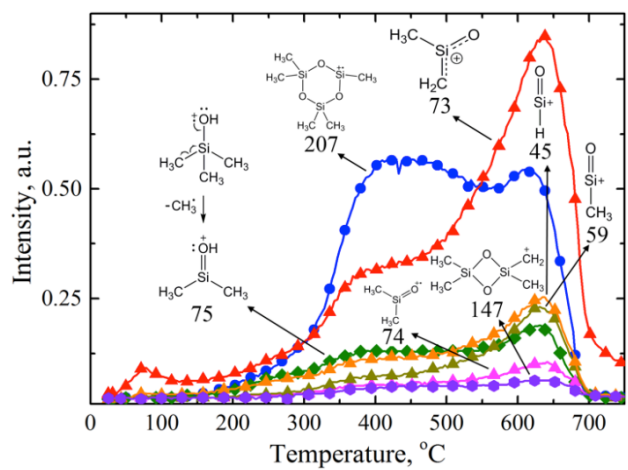

**Figure 5.** TPD MS curves for a PDMS/CeO$_2$/SiO$_2$_high sample. Numbers above the curves correspond to the *m/z* values.

**Explanation of nano-oxides catalytic activity**

The enhanced release of the dimethylsilanone from the surfaces of oxide samples in comparison to the pure polymer, and the fact that an increase in the ceria content catalyzes this process, can be rationalized by considering the surface chemistry of oxides and the presumable reaction mechanism. The catalytic behavior of nano-silica and nano-oxide mixtures CeO$_2$/SiO$_2$ is consistent with the mechanism proposed in Scheme 2 (Pathway 1) for the reaction of dimethylsilanone elimination.

Our TPD data shows that dimethylsilanone is generated even during thermolysis of pure PDMS and that nano-oxides introduced into the polymer shift the process to lower temperatures and intensify the dimethylsilanone generation stage at $T_{max}$ = ~ 630 – 650 °C in the following order: SiO$_2$ < SiO$_2$/CeO$_2$_low < SiO$_2$/CeO$_2$_high (relatively to the other products).

The surface silanol groups of silica are weakly acidic and are thus capable of forming hydrogen bonds.[63] This matches very well with the assumed reaction mechanism (Scheme 2, Pathway 1) and observed catalytic activity (Figures 3-5). Elimination of dimethylsilanone was proposed to proceed through a four-centered transition state by the intramolecular nucleophile attack of the siloxane oxygen on the silicon atom located next to the nearest [–SiO(CH$_3$)$_2$–] monomeric unit. The catalytic action of the silanols of nano-silica involves hydrogen bonding to the siloxane oxygen atom leading to the polarization of the Si-O bond, an increase in the Si atom's electrophilicity and, as a result, promotes dimethylsilanone elimination according to Scheme 3a.

Addition of ceria nanoparticles to the polymer/oxide composition brings Lewis acidity to the system.[64] The density of Lewis acidic sites increases with increasing ceria content. This again matches very well with the increased intensity of dimethylsilanone desorption and the proposed reaction mechanism. An incompletely coordinated atom of cerium on the surface of ceria nanoparticle acts as a Lewis acid site by coordinating to the oxygen atom of the polymer siloxane bond and withdrawing the electron density, which in turn facilitates heterolytic Si-O bond cleavage, promotes further Si=O double bond formation, and activates another siloxane oxygen for nucleophilic attack (Scheme 3b). Both silica and ceria/silica nano-oxides can also catalyze the reactions depicted in Scheme 2 (Pathway 2 and 3) by analogous mechanisms. These mechanistic suggestions are supported by the calculated kinetic parameters presented in the subsection below.

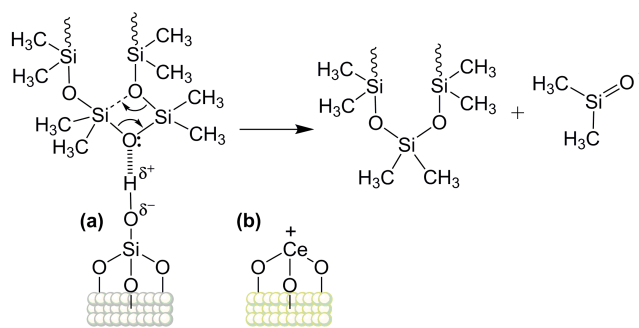

**Scheme 3.** Catalytic action of (a) surface silanols of silica and (b) Lewis acidic sites of ceria nanoparticles.

While the TPD MS technique is not inherently quantitative in an absolute sense, it is still possible to estimate the fraction of dimethylsilanone generated with respect to the other products by comparing the areas under the TPD curves for the ions of interest. According to the current literature, under vacuum conditions, thermal degradation of PDMS primarily results in depolymerization and formation of cyclic oligomers as products.[12-15] The trimer – hexamethylcyclotrisiloxane (HMCTS) is known to be the most abundant compound among these products. That is why to relatively assess the fraction of dimethylsilanone evolved we compared the amount of released dimethylsilanone (*m/z* = 73) to the amount of the released HMCTS (*m/z* = 207).

The TPD curves for the ions with *m/z* = 73 and 207 were analyzed by means of a multiple peak fitting procedure, where Gaussian functions were used to fit all the temperature maxima on the TPD curves of both dimethylsilanone and HMCTS. $R^2$ values were found to lie in the acceptable range (≥ 0.95) for all the plots. The total areas under the TPD curves for dimethylsilanone and HMCTS were obtained and their ratios were compared. Following values of the ratio (TPD curve area for *m/z* 207 : TPD curve area for *m/z* 73) were obtained: 0.83 – for pure PDMS; 0.97 – for PDMS/SiO$_2$; 0.97 – for PDMS/CeO$_2$/SiO$_2$_low; 0.94 – for PDMS/CeO$_2$/SiO$_2$_high.

Therefore, this analysis shows that dimethylsilanone is, just as the HMCTS, a major product of PDMS pyrolysis. The amount of the released dimethylsilanone is commeasurable with the amount of released HMCTS (please see Supporting Information for details). The highest relative quantity of dimethylsilanone was released from the pure PDMS sample. Samples of PDMS filled with $SiO_2$; $CeO_2/SiO_2$_low and $CeO_2/SiO_2$_high showed an increase (13-17 %) in the fraction of generated HMCTS in comparison to the generated dimethylsilanone. This can be explained by the fact that while nano-oxides catalyze both processes ($T_{max}$ for dimethylsilanone release shifts to lower values), the catalytic action is more prominent for the depolymerization reaction (HMCTS release stage at $T_{max}$ ~ 400 °C intensifies) than for dimethylsilanone formation. The efficient catalytic effect of silica and ceria nanoparticles on the PDMS thermal depolymerization has also been reported earlier.[12]

**Low-temperature desorption of dimethylsilanone**

The emergence of the low-temperature desorption maxima ($T_{max}$ = ~ 70 °C) for the samples of filled polymer is surprising (Figures 3-5). To investigate this further a three times larger quantity of the $PDMS/SiO_2$ sample was placed in the mass spectrometer (Figure 6). Analysis showed the release of dimethylsilanone ($m/z$ = 74) at $T_{max}$ = 68 ± 2.5 °C (3.6 % relative error), which is also confirmed by the release of its accompanying fragment ions ($m/z$ = 73, 59 and 45). Interestingly the simultaneous release of the TMCDS cation with $m/z$ = 147 (~ 6 % relative to $m/z$ = 73) and its parent ion $m/z$ = 148 (~ 1 % relative to $m/z$ = 73) was also registered at this temperature, whereas HMCTS was not observed. This indicates the validity of Pathways 1 and 3 (Scheme 2) as possible reaction mechanisms for dimethylsilanone formation at this temperature. The appearance of low-temperature desorption maximum is puzzling. However, contamination by impurities can be excluded since it was not observed for pure polymer pyrolysis. Presumably it could be due to the influence of some of the surface catalytic sites, but further investigations will be required before a firm conclusion can be made. Small differences in the $T_{max}$ values and distortions in the TPD curves shapes most likely originate from the interference of some other desorption products. In general these mismatches are not significant and can thus be neglected.

Interestingly, the kinetic parameters for the low temperature release of dimethylsilanone ($m/z$ = 74) and TMCDS ($m/z$ = 147) are different, see Table 1. Moreover, they differ from the values obtained for the high temperature stage of dimethylsilanone and TMCDS desorption. The most prominent and intriguing difference is that at low temperature the desorption kinetics of dimethylsilanone and TMCDS is well described by both a second-order model ($R^2$ = 0.974) and a first-order model ($R^2$ = 0.958). This may indicate the possibility of the gas phase decomposition of TMCDS into two molecules of dimethylsilanone (Scheme 2, Pathway 3).

Such a process very likely changes the shape of TPD maxima as it is known that the shape of the desorption maximum depends on the order of the desorption process.[60] TMCDS decomposition could also lead to an increase in the intensity of the ions with $m/z$ = 74, 73, 59, 43 (dimethylsilanone pattern), and to a decrease in the intensity of the ion with $m/z$ = 147. As such more complex rate laws need to be used for the description of the kinetics of the observed low-temperature desorption maxima as it seems that at least three parallel reactions are occurring (dimethylsilanone desorption, TMCDS desorption, TMCDS decomposition). Correspondingly this could be influencing the calculated kinetic parameters. The formation of TMCDS as a result of the condensation of two dimethylsilanone molecules is unlikely to proceed in vacuum. However, some cyclizations of diorganylsilanones to form tetraorganylsiloxanes have been reported previously.[19,65] Tetraorganylcyclodisiloxanes are known to be extremely unstable compounds that readily trap diorganylsilanones to form stable cyclotrisiloxanes.[19] However, we have not observed simultaneous formation of HMCTS at this temperature (Fig. 6).

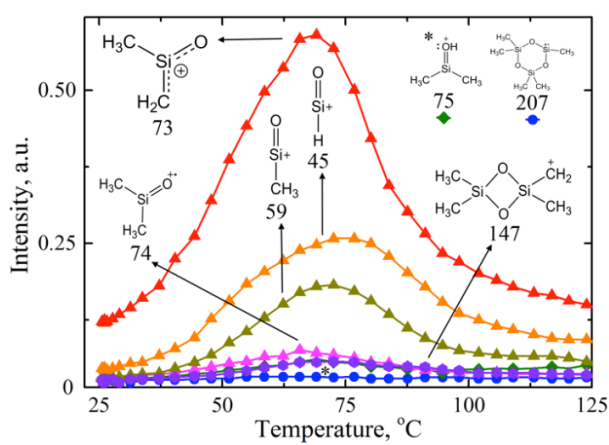

**Figure 6.** TPD MS curves for a high loading measurement of a PDMS/SiO$_2$ sample. Numbers above the curves correspond to the $m/z$ values.

**Kinetics of dimethylsilanone formation**

Kinetic parameters (the reaction order $n$, the activation energy $E^{\neq}$, the pre-exponential factor $v_0$, and activation entropy $dS^{\neq}$) of the dimethylsilanone formation reaction were determined from the TPD MS data by using the Arrhenius plot method. The detailed procedure for obtaining the kinetic parameters has been described earlier.[12] We were able to obtain the kinetics only for the high-temperature stage of the dimethylsilanone release for the samples PDMS, PDMS/CeO$_2$/SiO$_2$_low, and PDMS/CeO$_2$/SiO$_2$_high (as the other desorption maxima for these samples have ambiguous forms, some of the peaks overlap), see Table 1. The analysis of the kinetics data yielded higher R squared values and lower Standard Error of the Regression for the first-order model ($R^2$ = 0.953 - 0.975) than for the second-order one ($R^2$ = 0.851 - 0.929), see Table 1. It can be concluded that the dimethylsilanone formation most likely proceeds as a first order reaction via a highly ordered transition state (Scheme 3), which is consistent with the negative values of the entropy of activation (Table 1). This contradicts with the suggestion by Lewicki et al.[15] that elimination of dimethylsilanone at high temperatures in the PDMS/montmorillonite systems proceeds via radical scission mechanism. Our data is in a good agreement with those summarized in the review by Voronkov – where the chain rupture leading to dimethylsilanone elimination is believed to occur via highly ordered four-centered transition state.[19] The fact that the Si-O bond is very stable towards homolytic cleavage as well as towards splitting into biradical species also supports our interpretation. Earlier, however, we have shown that the other high-temperature processes occurring during PDMS/nano-oxide pyrolysis most likely proceed via radical mechanisms – formation of methane and ethylene.[12]

## Conclusions

Desorption mass spectrometry was successfully used to detect dimethylsilanone generation over a broad temperature range from pure polydimethylsiloxane polymer and its compositions with nanosized silica and ceria/silica. The mechanism of dimethylsilanone fragmentation under the influence of electron ionization was discussed in view of density functional theory calculations of the dissociation energies for different decay pathways. TPD curves of the dimethylsilanone molecular and fragment ions were found to reflect the shapes of each other, which supports the idea of them arising from the same process. Analysis of the TPD profiles suggests three different channels to be responsible for dimethylsilanone generation: 1) directly from PDMS polymer chains; 2) via gas-phase elimination from HMCTS; 3) from decomposition of desorbed TMCDS. Calculated kinetic parameters suggest that the dimethylsilanone formation reaction proceeds as a first order reaction via highly ordered transition state. Molecular and fragment ions corresponding to another elusive compound – TMCDS, were also observed.

Finally, it should be mentioned that the origin and the kinetics of the low-temperature (~ 70 °C) release of dimethylsilanone from PDMS/nano-oxide compositions is puzzling and requires further studies (we are considering combined TPD MS/chemical trapping experiments). However, the detection of this reactive compound being generated over a broad temperature range (up to 700 °C) extends the general understanding of the thermal decomposition chemistry of the PDMS-based nanocomposites.

**Acknowledgements**

The authors gratefully acknowledge the financial support from the Swedish Research Council in the framework of the Swedish Research Links programme grant 348-2014-4250 and Contract No. 621-2012-3660.

**Keywords:** dimethylsilanone • polydimethylsiloxane • pyrolysis • nanocomposite • silica